# Long-term prediction intervals of economic time series

M. Chudý[1,2]  · S. Karmakar[3] · W. B. Wu[4]

**Abstract**
We construct long-term prediction intervals for time-aggregated future values of univariate economic time series. We propose computational adjustments of the existing methods to improve coverage probability under a small sample constraint. A pseudo-out-of-sample evaluation shows that our methods perform at least as well as selected alternative methods based on model-implied Bayesian approaches and bootstrapping. Our most successful method yields prediction intervals for eight macroeconomic indicators over a horizon spanning several decades.

**JEL Classification** C14 · C15 · C53 · C87 · E27

## 1 Introduction

Long-term predictions of economic time series are published yearly by the US Congressional Budget Office (CBO) in its Long-term Budget and Economic Outlook.[1] In this report, the CBO predicts US federal spending and revenue growth in the coming decades under the assumption of stable tax and spending policies. However, structural changes occur over the long run (taking the turbulent period after the Great Moderation as an example), and not only as a result of changes in legislation. The CBO stated in its

---

[1] Available from https://www.cbo.gov/publication/52480.

✉ M. Chudý
    marek.chudy@univie.ac.at

1   Institute for Financial Policy, Ministry of Finance, Stefanovicova 5, 817 82 Bratislava, Slovak Republic
2   Department of Statistics and Operations Research, University of Vienna, Oskar Morgenstern Platz 1, 1090 Vienna, Austria
3   Department of Statistics, University of Florida, 230 Newell Drive, Gainesville, FL 32611, USA
4   Department of Statistics, University of Chicago, 5747 S. Ellis Avenue, Chicago, IL 60637, USA

January 2000 Budget and Economic Outlook that the baseline projections allow for an average recession within the next 10 years (2000–2010). Today, we know that the 2008 recession was more severe than the predicted average recession. Moreover, in its 2011 report[2] the US Financial Crisis Inquiry Commission concluded that the crisis would have been avoidable if timely preventive measures had been introduced. We do not link the absence of these measures with the CBO's projections from 18 years ago, but we do believe that accurate long-term economic predictions can trigger right and timely decisions. Economic predictions for several decades ahead are crucial for strategic decisions made by trust funds, pension management and insurance companies, portfolio management of specific derivatives (Kitsul and Wright 2013) and assets (see Bansal et al. 2016). Several facts hamper the long-term prediction of economic time series: small sample size because most post-WWII economic indicators are reported on monthly/quarterly bases, (anti-) persistence[3] (see Diebold and Rudebusch 1989; Baillie 1996; Diebold and Linder 1996, who also give PIs), heteroscedasticity and structural change (Cheng et al. 2016), the latter of which is inevitable in the long run (Stock and Watson 2005).

Sometimes, decision makers call for predictions of boundaries $[L, U]$ covering the future value of interest with a certain probability. Unlike point forecasts, prediction intervals (PIs) can capture the uncertainty surrounding predictions. As a proxy for this uncertainty, one can look at the widths of different PIs. Most software packages offer PIs as part of their standard output. PIs from exponential smoothing, for instance, are readily available without any strict assumptions, but then as the forecast horizon grows, these PIs often become too wide to be informative (see Chatfield 1993, for more background). By contrast, PIs implied by *arma–garch* models often turn out to be too narrow because they ignore distributional, parameter and model uncertainty (see Pastor and Stambaugh 2012). Pascual et al. (2004, 2006) therefore compute predictive densities using bootstrapping without the usual distributional assumptions while incorporating parameter uncertainty. Using Bayesian methods, one can account for both model and parameter uncertainty, but the pre-assigned coverage of PIs is attained only on average relative to the specified prior. Müller and Watson (2016) construct bayes PIs for temporal averages of economic series' growth rates over a horizon of 10–75 years. Using the so-called least favorable distribution solves the problem above with the pre-assigned coverage and makes their PIs more conservative. Zhou et al. (2010) provide theoretically valid long-term PIs for the same type of target as Müller and Watson (2016), i.e., the temporal aggregate of series over a long horizon. As opposed to Müller and Watson (2016), Zhou et al. (2010) do not require any model fitting (at least in our univariate setup) and thus provide a very simple alternative. While both papers allow for the presence of a long-memory component in the data generating process (DGP), Zhou et al. (2010) did not apply their methods to economic time series nor has either of the papers compared themselves with any benchmark. These facts pave the way for the following empirical research:

– First, since Zhou et al. (2010) evaluate their PIs using only simulated data, we find it necessary to verify their results using real data.

---

[2] Available from https://www.govinfo.gov/app/details/GPO-FCIC.

[3] Anti-persistence can be observed as well, often as a result of (over-) differencing.

- Second, the methods of Zhou et al. (2010), although theoretically valid, do not to account for some characteristics of economic time series. Therefore, we propose computational adjustments of the PIs of Zhou et al. (2010) that lead to better predictive performance for small samples and long horizons. Our adjustments employ a stationary bootstrap (Politis and Romano 1994) and kernel quantile estimators (Sheather and Marron 1990).
- Third, since neither Zhou et al. (2010) nor Müller and Watson (2016) compare their PIs to any benchmark, we take over this responsibility and conduct an extensive pseudo-out-of-sample (POOS) comparison. We augment the comparison with PIs implied by *arfima–garch* models computed as one of the following: (i) forecasts for time-aggregated series or (ii) time-aggregated forecasts of disaggregated series. To compute (i) and (ii) we use both analytic formulas and bootstrap path simulations (Pascual et al. 2006).

The main results of our paper may be summarized as follows:

- First, our simulation study reveals that the PIs of Zhou et al. (2010) fail to achieve their nominal coverage rate under a growing horizon as a result of rapidly shrinking width. Particularly under long-memory DGP, the coverage rate reaches only half of the nominal level.
- Second, using the proposed computational adjustments, we achieved an improvement in the coverage rate of 20pp, which may, however, still be below the nominal level.
- Third, based on real data (S&P 500 returns and US 3-month TB interest rates), the adjusted PIs of Zhou et al. (2010) provide a valid competitor for Müller and Watson (2016). Particularly in case of asset returns, the PIs of Müller and Watson (2016) provide higher coverage but less precision (larger width), while for assets' volatility, the roles are switched. In both cases, adjusted Zhou et al. (2010) PIs outperform the bootstrap PIs of Pascual et al. (2006).
- Fourth, with the adjusted method of Zhou et al. (2010), we construct long-term prediction intervals for selected US macroeconomic time series including GDP growth, total factor productivity, inflation, population, and others. These PIs provide an alternative for PIs given by Müller and Watson (2016) in Table 5 on pages 1731–1732 in the referenced paper.

Our article is organized as follows: In Sect. 2 we review the methods discussed above with a focus on their scope and implementation. We further describe our computational adjustments of both methods of Zhou et al. (2010) and justify them using simulations. Section 3 summarizes the empirical comparison of all previously discussed methods. Section 4 provides PIs for eight macro-indicators over the horizon of up to seven decades from now. Section 5 contains concluding remarks. Plots and details concerning implementation and underlying theory are available in "Appendix."

## 2 Methods and simulations

In this section, we first briefly discuss the three selected approaches for computing of PIs followed by their merits and demerits. Then we propose computational adjustments of the methods proposed by Zhou et al. (2010). Next, our simulations show that these adjustments improve the coverage when the horizon $m$ is large compared to the sample size $T$, for example when $m = T/2$. In the following, assume that we observe $y_1, \ldots, y_T$ and we want to provide a PI for the temporal average $(y_{T+1} + \cdots + y_{T+m})/m$. For the rest of the paper, we use the following notation (and analogous for the process of innovations $e_t$)

$$\bar{y} = \frac{1}{T} \sum_{t=1}^{T} y_t, \quad \bar{y}_{+1:m} = \frac{1}{m} \sum_{t=T+1}^{T+m} y_t, \quad \bar{y}_{t(m)} = \frac{1}{m} \sum_{j=1}^{m} y_{t-j+1}. \quad (2.1)$$

### 2.1 Methods for computing prediction intervals of temporal averages

#### 2.1.1 Bootstrap prediction intervals by Pascual et al. (2004, 2006)

For a specific description of their approach, let us assume a weakly stationary *arma*(1,1)–*garch*(1,1) process of the form

$$y_t = \phi y_{t-1} + e_t + \theta e_{t-1}, \quad e_t = \sigma_t \varepsilon_t, \quad \varepsilon_t \sim WN, \quad \sigma_t^2 = \omega + \alpha e_{t-1}^2 + \beta \sigma_{t-1}^2. \quad (2.2)$$

In order to obtain PIs for $y_{T+m}$, one typically uses the estimated MSE predictors of $y_{T+m}$ and $\sigma_{T+m}$ given the past observations

$$\hat{y}_{T,T+m} = \hat{\phi}^m y_T + \hat{\phi}^{m-1} \hat{\theta} \hat{e}_T, \quad (2.3)$$

$$\hat{\sigma}_{T,T+m}^2 = \frac{\hat{\omega}}{1 - \hat{\alpha} - \hat{\beta}} + \left(\hat{\alpha} + \hat{\beta}\right)^{m-1} \left(\hat{\sigma}_{T+1}^2 - \frac{\hat{\omega}}{1 - \hat{\alpha} - \hat{\beta}}\right). \quad (2.4)$$

The resulting analytic (anlt) PIs have the form

$$[L, U] = \hat{y}_{T,T+m} + [Q^N(\alpha/2), Q^N(1-\alpha/2)] \left(\sum_{j=1}^{m} \hat{\sigma}_{T,T+j}^2 \hat{\Psi}_{m-j}^2\right)^{1/2}, \quad (2.5)$$

where $\hat{\Psi}_0 = 1$ and $\hat{\Psi}_j, j = 1, \ldots, m-1$ are the estimates of coefficients from the causal representation of $y_t$. $Q^N$ denotes normal quantile. Besides the fact that these PIs ignore the parameter uncertainty, they would be inappropriate for heavy-tailed processes or when the innovations distribution is asymmetric. In order to deal with these issues, Pascual et al. (2004) introduce a re-sampling strategy using the estimated innovations in order to simulate $y_t, t = T+1, \ldots, T+m$, and then compute the conditional distribution of $y_{T+m}$ directly, avoiding strict distributional assumptions on the innovations. Their approach does not need a backward representation and thus captures *garch*-type processes. They also show the validity of their PIs for *arima* processes. However, we are not aware of any extension of these results for *arfima*. Computationally, it is simple to obtain both the analytic and bootstrap PIs implied by *arfima* models, since all necessary ingredients are readily available in rugarch R-package (see Ghalanos 2017) which efficiently implements the bootstrap PIs of Pascual et al. (2004, 2006). While the re-sampling can also incorporate parameter uncertainty, Pascual et al. (2006) show that for the *garch* models, coverage of PIs is similar whether one accounts for parameter uncertainty or not. The superiority of bootstrap PIs over the analytic PIs (2.5) prevails especially when the innovations distribution is asymmetric.

In order to obtain PIs for $\bar{y}_{+1:m}$ with *arfima–garch*-type models, we can either (i) use averages of the in-sample observations $\bar{y}_{t(m)}$, as defined above, or (ii) average the forecasts of $y_t$, over $t = T + 1, \ldots, T + m$. In both cases, we fit *arfima*$(p, d, q)$–*garch*$(P, Q)$ models to the data with the rugarch R-package. As already mentioned, the full re-sampling scheme takes into account the parameter uncertainty, however, for minor improvement of performance and high cost concerning computation time. Therefore, we use a partial re-sampling scheme which accounts for the uncertainty due to the unknown distribution of innovations. The fractional parameter $d \in [0, 0.5)$ is, depending on the series, either fixed to 0 (only for stock returns, see Sect. 3) or estimated by maximum likelihood (ML). The *arma* orders are restricted to $p, q \in \{1, \ldots, 4\}$ and are selected by *aic*. The *garch* orders are restricted to $(P, Q) \in \{(0, 0), (1, 1)\}$. The details of our implementation follow:

*Fitting arfima–garch to averaged in-sample observations (avg-series):*
1. Compute the series of overlapping rolling averages $\bar{y}_{t(m)} = m^{-1} \sum_{i=1}^{m} y_{t-i+1}$, for $t = m, \ldots, T$.
2. Fit the selected *arfima–garch* model to the series of $\bar{y}_{t(m)}$.
3. Compute

    (anlt) $m$-step-ahead MSE forecasts $\hat{\bar{y}}_{T,+1:m}$ and $\hat{\bar{\sigma}}^2_{T,+1:m}$ analogously to (2.3) by substituting the observations $y_t$ by rolling averages $\bar{y}_{t(m)}$. Then, PIs are given by (2.5).[4]

    (boot) residuals $\hat{e}_t, t = 1, \ldots, T$, and generate $b = 1, \ldots, B$ future paths $\hat{\bar{y}}^b_{T(m),t}, t = T + 1, \ldots, T + m$, recursively using (2.5) and the parameter estimates from the original sample. Obtain the PIs by inverting the empirical distribution of $\hat{\bar{y}}^b_{T(m),T+m}, b = 1, \ldots, B$.

*Fitting arfima–garch to original series and averaging forecasts (avg-forecasts):*
1. Fit the selected *arfima–garch* model to the series $y_t$.
2. Compute

    (anlt) $\hat{\bar{y}}_{T,+1:m} = m^{-1} \sum_{i=1}^{m} \hat{y}_{T,T+i}$, with $\hat{y}_{T,T+i}$ the $i$-step-ahead analytic forecast from (2.3). The scaling factor in PI $[L, U] = \hat{\bar{y}}_{T,+1:m} + [Q_t(\alpha/2), Q_t(1 - \alpha/2)]\hat{as}_{T,+1:m}$ is derived in "Appendix C."

    (boot) residuals $\hat{e}_t, t = 1, \ldots, T$ and generate $b = 1, \ldots, B$ future paths $\hat{y}^b_{T,t}, t = T+1, \ldots, T+m$, recursively using (2.5) and the parameter estimates from the original sample. Compute the temporal averages $\hat{\bar{y}}^b_{T,+1:m}$, as estimators of $\bar{y}^b_{T,+1:m}, b = 1, \ldots, B$. We obtain the PIs by inverting the empirical distribution of $\hat{\bar{y}}^b_{T,+1:m}, b = 1, \ldots, B$.

---

[4] Instead of the normal quantiles, we rather utilize Student's t where df is estimated by ML.

### 2.1.2 Robust bayes prediction intervals by Müller and Watson (2016)

With both the sample size and horizon growing proportionally, Müller and Watson (2016) provide asymptotically valid long-term PIs under a rich class of models for series with long memory under a unified spectral representation. In order to capture a larger scope of long-run dynamics in economic time series beyond those described by *arfima* models, Müller and Watson (2016) consider two additional models, namely (i) the *local-level* model

$$y_t = y_{1t} + (bT)^{-1} \sum_{s=1}^{t} y_{2s},$$
$$\text{where } \{y_{1t}\}, \{y_{2t}\} \text{ are mutually independent I(0) processes,} \quad (2.6)$$

and (ii) the *local to unity ar(1)* model defined by:

$$y_t = (1 - c/T)y_{t-1} + y_{1t}, \quad \text{where } \{y_{1t}\} \text{ is an I(0) process.} \quad (2.7)$$

The former captures varying "local means" arising, e.g., from stochastic breaks, while the latter is useful for modeling highly persistent series. In (i) the role of the persistent component is determined by the parameter $b$, while in (ii) it is driven by $c$. The *arfima* models with fractional integration parameter $d$ complete the triple of models in Müller and Watson (2016) who design a unified spectral representation of their long-run dynamics using the parametrization $\vartheta = (b, c, d)$.

A natural way of how to incorporate the uncertainty about $\vartheta$, which is crucial for the asymptotic predictive distribution of $\bar{y}_{+1:m}$, is to assume a prior for $\vartheta$. A practical drawback of such an approach is that the pre-assigned coverage holds only on average relative to the prior. Hence, Müller and Watson (2016) further robustify their bayes PIs in order to attain "frequentist coverage," i.e., coverage that holds over the whole parameter space.

The main idea behind their approach is to extract the long-run information from selected low-frequency projections of $y_t$, $t = 1, \ldots, T$. Assume that the set of predictors for $\bar{y}_{+1:m}$ consists of $q$ low-frequency cosine transformations $X = (X_1, \ldots, X_q)^\top$ of $y_t$. Then the asymptotic conditional density of $\bar{y}_{+1:m}$ is a function of the covariance matrix of $(X_1, \ldots, X_q, \bar{y}_{+1:m})$ denoted as $\Sigma$, which in turn can be expressed as a function of properly scaled spectra $S(m/T, q, \vartheta)$. When the number of frequencies $q$ is kept small, the high-frequency noise is filtered out, thus providing more robustness. For fixed $\vartheta = (0, 0, 0)$ the conditional distribution of $\bar{y}_{+1:m}$ turns out to be Student's t with $q$ degrees of freedom and the PIs take the form

$$[L, U] = \bar{y} + [Q_q^t(\alpha/2), Q_q^t(1 - \alpha/2)]\sqrt{\frac{m+T}{mq}X^\top X}. \quad (2.8)$$

These (*naive*) PIs implied by fixed $\vartheta = (0, 0, 0)$ can be enhanced by imposing a uniform prior on $\vartheta$, giving equal weight to all combinations of parameters $-0.4 \leq d \leq 1$, $b, c \geq 0$, and using standard Bayesian procedure to obtain posterior predictive density, which is no longer a simple $t$-distribution but rather a mixture of different $t$-distributions. We denote the implied PIs as (*bayes*) *PIs*. Finally, the (*robust*) PIs additionally guarantee the correct coverage uniformly across the parameter space $\Theta$ and simultaneously have optimal (mean) width. We conclude by giving our implementation steps for the (*bayes*) PIs, leaving the additional steps necessary for computing the (*robust*) PIs to our "Appendix B."

*Bayes PIs (bayes):*

1. Set $q$ small and compute the cosine transformations $X = (X_1, , X_q)$ of the target series $y_t$. Standardize them as $Z = (Z_1, \cdots, , Z_q)^\top = X/\sqrt{X^\top X}$.
2. For a chosen grid of parameter values $\vartheta = (b, c, d)$ satisfying $-0.4 \leq d \leq 1$; $b, c \geq 0$ compute the matrices $\Sigma(m/T, q, \vartheta)$ following formulas (9) and (20) from Müller and Watson (2016) and using, e.g., a numerical integration algorithm. (Details are given in "Appendix" of the original paper.)
3. Choose a prior for $\vartheta = (b, c, d)$ and compute the posterior covariance matrix $\Sigma = \begin{pmatrix} \Sigma_{ZZ} & \Sigma_{Z\bar{e}} \\ \Sigma_{Z\bar{e}} & \Sigma_{\bar{e}\bar{e}} \end{pmatrix}$.
4. Obtain the covariance matrix of the residuals as $\Sigma_{UU} = \Sigma_{\bar{e}\bar{e}} - \Sigma'_{Z\bar{e}}(\Sigma_{ZZ}^{-1})\Sigma_{Z\bar{e}}$.
5. Compute the quantiles $Q_q^{\text{tmix}}(\alpha/2), Q_q^{\text{tmix}}(1-\alpha/2)$ of the conditional (mixture-t) distribution of $\bar{y}_{+1:m}$ using, e.g., sequential bisection approximation. (Details are given in "Appendix" of the original paper.)
6. The PIs are given by $[L, U] = \bar{y} + [Q_q^{\text{tmix}}(\alpha/2), Q_q^{\text{tmix}}(1-\alpha/2)]\sqrt{X^\top X}$.

### 2.1.3 Prediction intervals by Zhou et al. (2010)

For presentational clarity of their approach, assume

$$y_t = \mu + e_t, \tag{2.9}$$

where $e_t$ is a mean-zero stationary process and $\mu$ is the unknown deterministic mean. The PI for $y_t$ process will be constructed via that of the $\hat{e}_t = y_t - \hat{\mu}$ process by adding the $\hat{\mu} = \bar{y}$ to both components of the intervals. It is common practice and can also be proved to have the correct coverage using standard arguments. We first provide a summary of the two methods proposed in Zhou et al. (2010) and then we discuss their consistency.

CLT method (*clt*): If the process $e_t$ shows short-range dependence and light-tailed behavior, then in the light of a quenched CLT, Zhou et al. (2010) propose the following PI for $\bar{e}_{+1:m}$

$$[L, U] = [Q^N(\alpha/2), Q^N(1-\alpha/2)]\frac{\sigma}{\sqrt{m}}, \tag{2.10}$$

where $\sigma$ is the long-run standard deviation (sd) of $e_t$. However, since $\sigma$ is unknown, it must be estimated. One popular choice is the lag window estimator

$$\hat{\sigma}^2 = \sum_{k=-k_T}^{k_T} \hat{\gamma}_k = \sum_{k=-k_T}^{k_T} \frac{1}{T} \sum_{t=1}^{T-|k|} (\hat{e}_t - \bar{\hat{e}})(\hat{e}_{t+k} - \bar{\hat{e}}). \quad (2.11)$$

The PI for $\bar{y}_{+1:m}$ with nominal coverage $100(1-\alpha)\%$ is given by

$$[L, U] = \bar{y} + [Q^N(\alpha/2), Q^N(1-\alpha/2)]\frac{\hat{\sigma}}{\sqrt{m}}. \quad (2.12)$$

Quantile method (*qtl*): If we allow for heavy tails and long memory, the PI for $\bar{y}_{+1:m}$ with nominal coverage $100(1-\alpha)\%$ can be obtained by

$$[L, U] = \bar{y} + [\hat{Q}(\alpha/2), \hat{Q}(1-\alpha/2)], \quad (2.13)$$

where $\hat{Q}(\cdot)$ is the respective empirical quantile of $\bar{\hat{e}}_{t(m)}, t = m, \ldots, T$.

*The clt method* is applicable only for processes with light-tailed behavior and short-range dependence. Let $S_t = \sum_{1 \leq j \leq t} e_j$. Under stationarity, the problem of predicting $\bar{e}_{+1:m} = (S_{T+m} - S_T)/m$ after observing $e_1, \ldots, e_T$ can be analogically thought of as predicting $S_m/\sqrt{m}$ after observing $\ldots, e_{-1}, e_0$. Let $\mathcal{F}_0$ be the $\sigma$-field generated by $\ldots, e_{-1}, e_0$. Assume $\mathbb{E}(|e_t|^p) < \infty$ for some $p > 2$. Wu and Woodroofe (2004) proved that, if for some $q > 5/2$,

$$\|E(S_m|\mathcal{F}_0)\| = O\left(\frac{\sqrt{m}}{\log^q m}\right), \quad (2.14)$$

then we have the a.s. convergence

$$\Delta(\mathbb{P}(S_m/\sqrt{m} \leq \cdot|\mathcal{F}_0), N(0, \sigma^2)) = 0 \text{ a.s.}, \quad (2.15)$$

where $\Delta$ denotes the Levy distance, $m \to \infty$, and $\sigma^2 = \lim_{m \to \infty} \|S_m\|^2/m$ is the long-run variance. Verifying (2.14) is elementary for many well-known time series models. We postpone the discussion on the verification of such a result for a linear and nonlinear process for the interested reader to "Appendix D."

*The qtl method* is based on the intuitive fact that for the horizon $m$ growing to $\infty$ and in the light of weak dependence,

$$\mathbb{P}\left(a \leq \frac{e_{T+1} + \cdots e_{T+m}}{m} \leq b|e_1, \ldots, e_T\right) \approx \mathbb{P}\left(a \leq \frac{e_{T+1} + \cdots e_{T+m}}{m} \leq b\right), \quad (2.16)$$

if $m/T$ is not too small. Thus, it suffices to estimate the quantiles of $\bar{e}_{T(m)} = (e_{T+1} + \cdots + e_{T+m})/m$ using, e.g., empirical quantiles of $\bar{e}_{t(m)}, t = m, \ldots, T$. The power of this method lies in its applicability to heavy-tailed error process. Zhou et al. (2010) provided a consistency result for this method for the subclass of linear processes (see Theorem 2 in our "Appendix" section D).

### 2.1.4 Practical comparison of the previously discussed methods

*Pros and cons of* Pascual et al. (2004, 2006) When comparing the bootstrap PIs to the analytic PIs, the former provide the advantage of including the uncertainty due to the unknown distribution of the residuals and unknown parameters into the uncertainty about the target. In cases when the distribution of the residuals is asymmetric and doubts about the proximity of the estimated model to the true DGP exist, the bootstrap approach dominates the analytic. Furthermore, regarding the implementation, the analytic PIs are more difficult to obtain since we deal with a nonstandard target. By contrast, the bootstrap PIs are readily available in the R-package *rugarch*. Hence, their estimation is cheap. Concerning the two ways of fitting the models to data, i.e., (i) using the series of rolling temporal averages or (ii) using the original series and averaging the forecasts, there are pros and cons for each approach regarding the implementation and effective use of our relatively small sample. The literature (e.g., page 302 in Lütkepohl 2006; Marcellino 1999) does not provide any conclusion about the superiority of (i) over (ii), or vice versa. Therefore, we include both (i) and (ii) in the POOS comparison in Sect. 3.

*Pros and cons of* Müller and Watson (2016) Their methods represent the state of the art, being robust against stylized peculiarities of economic time series. Their Bayesian approach accounts for both model and parameter uncertainty, but the focus is only on those parameters ruling the persistence, which is in contrast to the previously discussed bootstrap approach where the focus is on the short-term dynamics. To date, no package implementation has been available, which makes the approach less attractive to practitioners. Moreover, the PIs depend on several forecaster-made choices, such as the number of frequencies $q$ to keep, the grid of values for parameters, the choice of prior. Even with these inputs fixed, the computation takes longer due to multiple advanced numerical approximations required for the (*bayes*) *PIs* and further optimization to attain the "frequentist coverage." PIs for fixed parameters $q = 12$ and $0.075 \leq m/T \leq 1.5$ used in their paper (and also in the current paper) are available faster thanks to some pre-computed inputs available from the replication files.[5]

*Pros and cons of* Zhou et al. (2010) Their methods provide a simple alternative to the previously discussed ones. As to their scope of applicability, the *clt* method does not require any specific rate of how fast the horizon can grow compared to the sample size. However, the predictive performance heavily depends on the estimator of the long-term volatility $\sigma$. Furthermore, for some processes with heavy-tailed innovations or long-range dependence, the notion of the long-run variance $\sigma^2$ does not exist, and thus this method is not applicable. The attractive feature of the *qtl* PIs is the simplicity and more general applicability than the *clt*. Their computation requires almost no optimization (at least in our univariate case) and is straightforward. Pascual et al. (2004, 2006) and Müller and Watson (2016) assume that the DGP of $y_t$ is (possibly long-memory and heteroscedastic version of) an *arma* process. Zhou et al. (2010) do not a priori assume any parametrization for the dynamics of $y_t$, but argue that both *qtl* and *clt* PIs are valid for *arma* processes, whereas only the former should be used for processes with a long memory. The simulations of Zhou et al. (2010) confirm their claims. However, as we demonstrate next, the *qtl* PIs under-perform when $T$ is small and $m/T \approx 1/2$. Therefore, we propose some computational adjustment and provide a simulation-based justification of their superiority over the original *clt* and *qtl*.

---

[5] The replication files for these methods are available in MATLAB from M. Watson's homepage.

### 2.2 Zhou et al. (2010) under small sample: adjustment and simulations

#### 2.2.1 Computational adjustments

The simulation setup in Zhou et al. (2010), page 1440, assumes $T = 6000$ and horizon $m = 168$. By contrast, in an economic forecasting setup, one typically has only a few hundred of observations, while our horizon $m$ stays approximately the same. Here we show how one can easily modify the computation of *clt* and *qtl* PIs in order to enhance their performance. In particular, for *qtl*, we use a stationary bootstrap (Politis and Romano 1994) with optimal window width as proposed by Politis and White (2004) and Patton et al. (2009) to obtain a set of replicated series. Next, kernel quantile estimators (see Silverman 1986; Sheather and Marron 1990) are used instead of sample quantiles. In order to improve the *clt* method, we employ a different estimator (cf. 2.18) of $\sigma$ than (2.11) and account for the estimation uncertainty. These three modifications are then shown to improve the empirical coverage using simulations.

*Stationary bootstrap* The procedure starts with the decomposition of the original sample into blocks by choosing the starting point $i$ and the length of block $L_i$ from a uniform and geometric distribution, respectively, that are independent of the data. For every starting point and length, we re-sample from the blocks of the original series. The resulting blocks are then concatenated. As proposed by Politis and Romano (1994) in their seminal paper, this way of re-sampling retains the weak stationarity and is less sensitive to the choice of block size than moving block bootstrap (Künsch 1989). It also retains the dependence structure asymptotically since every block contains consecutive elements of the original series. The two re-sampling schemes differ in the way how they deal with the end-effects. Under mixing conditions, the consistency of stationary bootstrap for the centered and normalized mean process has been studied in the literature. Gonçalves and de Jong (2003) show that under some mild moment conditions, for some suitable $c_T \to 0$,

$$\sup_x |\mathbb{P}^*(\sqrt{T}(\bar{y}^* - \bar{y}) \leq x) - \mathbb{P}(\sqrt{T}(\bar{y} - \mu) \leq x)| = o_\mathbb{P}(c_T) \qquad (2.17)$$

holds where $\bar{y}^*$ and $\mathbb{P}^*$ refer to the re-sampled mean and the probability measure induced by the bootstrap. We conjecture that, along the same line of proof shown by Zhou et al. (2010), it is easy to show consistency results for the bootstrapped versions of the rolling averages of $m$ consecutive realizations. This is immediate for linear processes. For nonlinear processes, one can use the functional dependence measure introduced by Wu (2005) and obtain analogous results. To keep our focus on empirical evaluations, we leave the proof of the consistency for our future work. Interested readers can also look at the arguments by Sun and Lahiri (2006) for moving block bootstrap and the corresponding changes as suggested in Lahiri (2013) to get an idea how to show quantile consistency result. For time series forecasting and quantile regression, the stationary bootstrap has been used by White (2000) and Han et al. (2016) among others.

*Kernel quantile estimation* The efficiency of kernel quantile estimators over the usual sample quantiles has been proved in Falk (1984) and was extended to several variants by Sheather and Marron (1990). As proposed in the latter, the improvement in MSE is a constant order of $\int u K(u) K^{-1}(u) du$ for the used symmetric kernel $K$. The theorems mentioned in Sect. 2 are easily extendable to these kernel quantile estimators. We conjecture that one can use the Bahadur-type representations for the kernel quantile estimators as shown in Falk (1985) and obtain similar results of consistency for at least linear processes. We used the popular Epanechnikov kernel $K(x) = 0.75(1-x^2)_+$ for our computations because of its efficiency in terms of mean integrated square error.

*Estimation of $\sigma$ and degrees of freedom* As mentioned above, Zhou et al. (2010) used *clt* as in (2.12) with normal quantiles. For many applications in economics and finance, the normal distribution fails to describe the possibly heavy-tailed behavior. Therefore, we propose to substitute the normal with the Student $t$-distribution, given the fact that $\sigma$ has to be estimated. Accounting for the estimation uncertainty indeed gives a Student $t$-distribution of $\bar{e}_{+1:m}$ in the limit. The question remains: How many degrees of freedom ($df$) we should use. Rather than some arbitrary choices such as 5, as used by default in many software packages, or ML-estimated df which would be very noisy given the small sample, we link them to the estimator of $\sigma$. This would not be trivial for the lag window estimator (2.11). Instead, we use the subsampling block estimator (see eq. 2, page 142 in Dehling et al. 2013)

$$\tilde{\sigma} = \frac{\sqrt{\pi l/2}}{T} \sum_{i=1}^{\kappa} \left| \sum_{t=(i-1)l+1}^{il} \hat{e}_t \right|, \quad (2.18)$$

with the block length $l$ and number of blocks $\kappa = \lceil T/l \rceil$. Then the adjusted *clt* $p = 100(1-\alpha)\%$ PI for $\bar{y}_{+1:m}$ is given by

$$[L, U] = \bar{y} + [Q^t_{\kappa-1}(\alpha/2), Q^t_{\kappa-1}(1-\alpha/2)] \frac{\tilde{\sigma}}{\sqrt{m}}, \quad (2.19)$$

where $Q^t_{\kappa-1}(\cdot)$ denotes a quantile of Student's t distribution with $\kappa - 1$ degrees of freedom. Note that, since we used non-overlapping blocks, under short-range dependence, these blocks behave almost independently and thus $\tilde{\sigma}^2$ with proper normalization constant behaves similarly to a $\chi^2$ distribution with $\kappa - 1$ degrees of freedom. Below, we give the implementation steps for the *adjusted qtl (kernel-boot)*:

1. Replicate series $e_t$, $B$ times obtaining $e_t^b, t = 1, \ldots, T, b = 1, \ldots, B$.
2. Compute $(\bar{e}^b_{t(m)}) = m^{-1} \sum_{i=1}^{m} e^b_{t-i+1}, t = m, \ldots T$ from every replicated series.
3. Estimate the $\alpha/2$th and $(1-\alpha/2)$th quantiles $\hat{Q}(\alpha/2)$ and $\hat{Q}(1-\alpha/2)$ using the Epanechnikov kernel density estimator from $\bar{e}^b_{T(m)}, b = 1, \ldots, B$.
4. The PI for $\bar{y}_{+1:m}$ is $[L, U] = \bar{y} + [\hat{Q}(\alpha/2), \hat{Q}(1-\alpha/2)]$.

Similarly, the implementation of the *adjusted clt method (clt-tdist):*

1. Estimate the long-run standard deviation from $e_t$, $t = 1, \ldots, T$, using the subsampling estimator (2.18) with block length as proposed by Carlstein (1986).
2. The PI is given by $[L, U] = \bar{y} + [Q^t_{\kappa-1}(\alpha/2), Q^t_{\kappa-1}(1 - \alpha/2)]\tilde{\sigma}/\sqrt{m}$.

### 2.2.2 Simulations

An extensive out-of-sample forecasting evaluation based on independent samples is possible only with artificial data. Our simulation setup is designed to assess the performance of the original methods of Zhou et al. (2010) as described in Sect. 2.1.3 and the computational modifications described in 2.2.1. The simulation results provide evidence for the usefulness of these modifications in an artificial setup based on possibly long-memory *arma*-like processes. This setup would provide an advantage for approaches described in 2.1.1 and 2.1.2, should we challenge them. We leave this task for the next section and real data.

We adopt the following four scenarios for the $e_t$ process from Zhou et al. (2010):

(i)  $e_t = 0.6 e_{t-1} + \sigma \epsilon_t$, for i.i.d mixture-normal $\epsilon_t \sim 0.5 N(0, 1) + 0.5 N(0, 1.25)$,
(ii) $e_t = \sigma \sum_{j=0}^{\infty}(j + 1)^{-0.8}\epsilon_{t-j}$, with noise as in (i),
(iii) $e_t = 0.6 e_{t-1} + \sigma \epsilon_t$, for stable $\epsilon_t$ with heavy tail index 1.5 and scale 1,
(iv) $e_t = \sigma \sum_{j=0}^{\infty}(j + 1)^{-0.8}\epsilon_{t-j}$, with noise as in (iii),

which correspond to (i) light tail and short memory, (ii) light tail and long memory, (iii) heavy tail and short memory, and (iv) heavy tail and long memory DGPs. For each scenario, we generate pseudo-data of length $T + m$, using the first $T$ observations for estimation and the last $m$ for evaluation. The experiment is repeated $N_{\text{trials}} = 10\,000$ times for each scenario.

The choice of parameters[6] $T = 260$, $m = 20, 30, 40, 60, 90, 130$ and $\sigma = 1.31$ mimics our setup for the real-data experiment in the next section. Following Müller and Watson (2016), we run our simulation for the nominal coverage probabilities $p = 1 - \alpha = 90\%$ (see Table 1A), resp. $= 67\%$ (see Table 1B), and compute the empirical coverage probability

$$\hat{p} = \frac{1}{N_{\text{trials}}} \sum_{i=1}^{N_{\text{trials}}} \mathbb{I}\left([L, U]_i \ni \bar{e}_{i,+1:m}\right), \quad (2.20)$$

where $\mathbb{I}$ for the $i$th trial is 1 when the future mean for the $i$th trial $\bar{e}_{i,+1:m}$ is covered by the $[L, U]_i$ and 0 otherwise. Furthermore, we report the relative median width

$$\hat{w} = \text{median}\left(|U - L|_1, \ldots, |U - L|_{N_{\text{trials}}}\right)/\left(\hat{Q}(1 - \alpha/2) - \hat{Q}(\alpha/2)\right), \quad (2.21)$$

where $\hat{Q}(\cdot)$ denotes the corresponding quantile of the empirical distribution of $\bar{e}_{+1:m}$, estimated from $\bar{e}_{i,+1:m}$, $i = 1, \ldots, N_{\text{trials}}$.

We focus on the evaluation under the longest horizon $m = 130$.

---

[6] The value of $\sigma$ was obtained from an $ar(1)$ model fitted to the full data set of S&P 500 returns.

**Table 1** Simulation results: comparison of coverage probabilities for two original methods proposed by Zhou et al. (2010) and the computational adjustments thereof proposed in Sect. 2.2.1

| | Horizon | Coverage probability $\hat{p}$ | | | | | Relative median width $\hat{w}$ | | | | | |
|---|---|---|---|---|---|---|---|---|---|---|---|---|
| | | 20 | 30 | 40 | 60 | 90 | 130 | 20 | 30 | 40 | 60 | 90 | 130 |

*(A) Results of simulated forecasting experiment for nominal coverage probability $p = 1 - \alpha = 90\%$*

| | | 20 | 30 | 40 | 60 | 90 | 130 | 20 | 30 | 40 | 60 | 90 | 130 |
|---|---|---|---|---|---|---|---|---|---|---|---|---|---|
| short-light | qtl-original | 85.25 | 82.33 | 78.85 | 72.85 | 62.19 | 47.97 | 0.94 | 0.90 | 0.86 | 0.77 | 0.64 | 0.47 |
| | qtl-kernel | 87.53 | 84.82 | 81.81 | 76.28 | 66.46 | 51.91 | 0.99 | 0.96 | 0.92 | 0.83 | 0.69 | 0.52 |
| | qtl-boot | 82.56 | 80.84 | 80.45 | 79.19 | 76.70 | 74.70 | 0.86 | 0.85 | 0.85 | 0.85 | 0.85 | 0.84 |
| | kernel-boot | 85.82 | 84.28 | 83.81 | 82.47 | 80.38 | 78.06 | 0.93 | 0.92 | 0.92 | 0.92 | 0.92 | 0.91 |
| | clt-original | 85.05 | 83.50 | 82.52 | 80.89 | 78.80 | 76.44 | 0.91 | 0.89 | 0.89 | 0.88 | 0.89 | 0.88 |
| | clt-tdist | 86.11 | 84.03 | 83.50 | 82.17 | 79.97 | 77.51 | 0.92 | 0.91 | 0.91 | 0.90 | 0.90 | 0.89 |
| long-light | qtl-original | 80.88 | 75.96 | 71.72 | 63.45 | 51.20 | 37.31 | 0.77 | 0.72 | 0.66 | 0.56 | 0.43 | 0.29 |
| | qtl-kernel | 83.45 | 79.06 | 74.99 | 67.89 | 56.40 | 41.39 | 0.82 | 0.76 | 0.71 | 0.61 | 0.48 | 0.32 |
| | qtl-boot | 76.42 | 72.20 | 69.25 | 64.21 | 58.54 | 53.78 | 0.68 | 0.63 | 0.59 | 0.53 | 0.48 | 0.44 |
| | kernel-boot | 80.58 | 76.02 | 73.25 | 68.43 | 62.87 | 57.63 | 0.74 | 0.69 | 0.64 | 0.58 | 0.52 | 0.48 |
| | clt-original | 77.33 | 70.98 | 66.96 | 61.12 | 55.01 | 50.18 | 0.68 | 0.60 | 0.55 | 0.49 | 0.44 | 0.40 |
| | clt-tdist | 84.44 | 78.50 | 74.80 | 68.74 | 63.26 | 57.96 | 0.82 | 0.72 | 0.66 | 0.58 | 0.53 | 0.48 |
| short-heavy | qtl-original | 84.39 | 80.21 | 76.71 | 70.12 | 58.54 | 44.45 | 0.99 | 0.87 | 0.78 | 0.66 | 0.51 | 0.36 |
| | qtl-kernel | 85.93 | 82.15 | 79.13 | 73.43 | 63.37 | 48.51 | 1.01 | 0.91 | 0.83 | 0.72 | 0.57 | 0.40 |
| | qtl-boot | 82.40 | 79.38 | 78.04 | 74.96 | 70.53 | 66.74 | 0.87 | 0.81 | 0.77 | 0.72 | 0.66 | 0.62 |
| | kernel-boot | 84.59 | 82.10 | 80.70 | 78.40 | 74.78 | 71.46 | 0.92 | 0.86 | 0.82 | 0.77 | 0.72 | 0.67 |
| | clt-original | 83.43 | 80.27 | 78.31 | 74.76 | 69.49 | 64.32 | 0.80 | 0.74 | 0.71 | 0.66 | 0.61 | 0.57 |
| | clt-tdist | 83.92 | 80.55 | 78.72 | 74.99 | 69.62 | 64.44 | 0.81 | 0.75 | 0.72 | 0.67 | 0.62 | 0.58 |

**Table 1** continued

| | | Coverage probability $\hat{p}$ | | | | | | Relative median width $\hat{w}$ | | | | | |
|---|---|---|---|---|---|---|---|---|---|---|---|---|---|
| | Horizon | 20 | 30 | 40 | 60 | 90 | 130 | 20 | 30 | 40 | 60 | 90 | 130 |
| long-heavy | qtl-original | 80.64 | 76.27 | 71.27 | 62.87 | 49.82 | 33.62 | 0.66 | 0.58 | 0.52 | 0.43 | 0.32 | 0.21 |
| | qtl-kernel | 82.53 | 78.51 | 74.32 | 66.80 | 55.13 | 37.49 | 0.69 | 0.62 | 0.56 | 0.47 | 0.36 | 0.24 |
| | qtl-boot | 78.23 | 74.59 | 70.16 | 64.31 | 57.06 | 50.64 | 0.58 | 0.52 | 0.47 | 0.41 | 0.35 | 0.31 |
| | kernel-boot | 80.96 | 77.22 | 73.38 | 68.14 | 61.08 | 54.23 | 0.62 | 0.55 | 0.50 | 0.44 | 0.38 | 0.34 |
| | clt-original | 77.73 | 71.64 | 66.70 | 59.51 | 51.70 | 44.34 | 0.54 | 0.46 | 0.41 | 0.35 | 0.29 | 0.26 |
| | clt-tdist | 83.24 | 78.34 | 73.81 | 67.49 | 60.01 | 52.46 | 0.67 | 0.57 | 0.50 | 0.43 | 0.37 | 0.32 |
| *(B) Results of simulated forecasting experiment for nominal coverage probability $p = 1 - \alpha = 67\%$* | | | | | | | | | | | | | |
| short-light | qtl-original | 62.79 | 60.68 | 59.19 | 54.49 | 45.88 | 33.48 | 0.97 | 0.92 | 0.91 | 0.85 | 0.71 | 0.52 |
| | qtl-kernel | 65.28 | 63.12 | 61.20 | 56.16 | 47.76 | 34.91 | 1.02 | 0.97 | 0.96 | 0.89 | 0.74 | 0.55 |
| | qtl-boot | 58.82 | 57.74 | 56.81 | 55.63 | 53.22 | 50.49 | 0.88 | 0.85 | 0.86 | 0.85 | 0.85 | 0.84 |
| | kernel-boot | 62.48 | 61.55 | 60.63 | 59.12 | 56.89 | 54.13 | 0.95 | 0.92 | 0.93 | 0.92 | 0.92 | 0.92 |
| | clt-original | 61.52 | 59.88 | 58.93 | 57.21 | 55.14 | 52.14 | 0.93 | 0.89 | 0.89 | 0.88 | 0.88 | 0.88 |
| | clt-tdist | 61.81 | 60.53 | 59.47 | 57.61 | 55.47 | 52.29 | 0.93 | 0.89 | 0.90 | 0.89 | 0.88 | 0.88 |
| long-light | qtl-original | 57.98 | 56.14 | 53.68 | 48.54 | 38.94 | 27.71 | 0.79 | 0.75 | 0.73 | 0.66 | 0.52 | 0.35 |
| | qtl-kernel | 60.30 | 58.05 | 55.88 | 49.70 | 40.39 | 28.75 | 0.83 | 0.79 | 0.76 | 0.68 | 0.54 | 0.36 |
| | qtl-boot | 52.78 | 50.01 | 47.72 | 43.92 | 38.84 | 35.42 | 0.69 | 0.64 | 0.60 | 0.55 | 0.49 | 0.45 |
| | kernel-boot | 56.27 | 53.11 | 50.60 | 46.86 | 41.46 | 37.82 | 0.74 | 0.69 | 0.65 | 0.59 | 0.53 | 0.48 |
| | clt-original | 53.22 | 47.78 | 44.50 | 39.71 | 34.97 | 31.64 | 0.68 | 0.60 | 0.55 | 0.49 | 0.44 | 0.40 |
| | clt-tdist | 59.29 | 54.20 | 50.21 | 45.16 | 40.08 | 36.17 | 0.79 | 0.69 | 0.64 | 0.57 | 0.51 | 0.46 |

**Table 1** continued

| | Horizon | Coverage probability $\hat{p}$ | | | | | | Relative median width $\hat{w}$ | | | | | | |
|---|---|---|---|---|---|---|---|---|---|---|---|---|---|---|
| | | 20 | 30 | 40 | 60 | 90 | 130 | 20 | 30 | 40 | 60 | 90 | 130 |
| short-heavy | qtl-original | 62.74 | 60.31 | 59.95 | 54.78 | 44.39 | 31.04 | 0.98 | 0.95 | 1.01 | 0.91 | 0.72 | 0.49 |
| | qtl-kernel | 65.36 | 63.16 | 61.99 | 56.50 | 46.58 | 32.43 | 1.04 | 1.02 | 1.04 | 0.95 | 0.76 | 0.52 |
| | qtl-boot | 60.15 | 57.80 | 57.52 | 55.54 | 51.06 | 46.67 | 0.91 | 0.89 | 0.90 | 0.87 | 0.81 | 0.77 |
| | kernel-boot | 63.89 | 61.67 | 60.90 | 58.78 | 54.62 | 50.24 | 0.99 | 0.97 | 0.97 | 0.94 | 0.88 | 0.84 |
| | clt-original | 64.49 | 59.31 | 56.29 | 51.00 | 44.63 | 40.45 | 1.00 | 0.91 | 0.87 | 0.81 | 0.75 | 0.71 |
| | clt-tdist | 65.12 | 59.64 | 56.94 | 51.68 | 45.07 | 40.71 | 1.01 | 0.92 | 0.88 | 0.82 | 0.75 | 0.71 |
| long-heavy | qtl-original | 59.02 | 55.99 | 54.33 | 48.08 | 37.93 | 24.88 | 0.74 | 0.70 | 0.69 | 0.61 | 0.47 | 0.31 |
| | qtl-kernel | 61.52 | 58.50 | 56.04 | 49.65 | 39.27 | 25.96 | 0.78 | 0.74 | 0.71 | 0.62 | 0.48 | 0.32 |
| | qtl-boot | 55.02 | 51.52 | 49.55 | 45.38 | 39.54 | 33.50 | 0.65 | 0.60 | 0.56 | 0.50 | 0.44 | 0.39 |
| | kernel-boot | 58.55 | 55.03 | 52.49 | 48.24 | 41.97 | 35.92 | 0.70 | 0.65 | 0.60 | 0.54 | 0.47 | 0.42 |
| | clt-original | 57.44 | 49.86 | 45.48 | 39.22 | 32.53 | 27.15 | 0.66 | 0.56 | 0.50 | 0.42 | 0.36 | 0.32 |
| | clt-tdist | 64.32 | 57.08 | 51.97 | 45.33 | 38.57 | 32.41 | 0.79 | 0.67 | 0.59 | 0.50 | 0.43 | 0.38 |

We simulate following scenarios: *short memory & light tail*, *long memory & light tail*, *short memory & heavy tail* and *long memory & heavy tail*. The reported values are coverage probabilities (in%) and median widths of the prediction intervals obtained for 10 000 out-of-sample trials. The median widths are reported relative to the width of the respective inter-quantile range computed from the empirical distribution of the targeted out-of-sample averages

*Scenario (i)* When $m = 130$ the original *qtl* covers the future realizations in only around 48% of cases, while the nominal coverage is 90%. Employing the kernel quantile adjustment on *qtl* increases this number by 4 percent points (pp), and when combined with the adjustment based on bootstrapping it yields an additional 26pp on top. Intuitively, using Student's t quantiles (instead of normal) leads to a higher coverage probability for the *clt*. As expected, the two methods perform similarly well in this particular scenario.

*Scenario (ii)* Long memory of the DGP has a strongly negative impact on both methods. The combined kernel-bootstrap adjustment increases the coverage of *qtl* by 20pp, which is, however, still very low. The same holds for the performance of *clt* under t-quantile adjustment.

*Scenario (iii)* Heavy-tailed noise has also a negative impact on the original *clt* (coverage probability falls by 13pp compared to the light-tailed case), whereas *qtl*, as expected, is more robust (falls by 4–6pp). The kernel-bootstrap adjustment increases coverage probability by 27pp for the *qtl*, whereas the *clt- tdist* yields only negligible improvement compared to the original *clt*.

*Scenario (iv)* The combined effect of (ii) and (iii) cuts the coverage probabilities further down—below 45%. The proposed adjustments increase the coverage probabilities by up to 10pp.

Overall, for the short and medium horizons, i.e., $m = 20, \ldots, 60$, we corroborate the conclusion from Zhou et al. (2010) that the (original) *clt* loses against the (original) *qtl*. However, both original methods exhibit rapid decay in their coverage probabilities as the forecasting horizon grows. For instance, in the scenario (iv) the gap between horizon $m = 20$ and $m = 130$ for the *qtl* is 47pp. Concerning the width of the PIs, we can see that both adjusted and original methods underestimate the dispersion and the gap between the width of PIs and the width of the empirical inter-quantile range increases with the horizon. However, our computational adjustments improve the original methods consistently over all scenarios. The improvement is most remarkable for the combined adjustment (*kernel-boot*).

## 3 Forecast comparison with long financial time series

This section summarizes a real-data POOS forecasting comparison for:

(zxw) adjusted PIs by Zhou et al. (2010),
(mw) robust bayes PIs by Müller and Watson (2016),
(prr) bootstrap PIs by Pascual et al. (2004, 2006) augmented by their analytical counterpart.

*Data and setup for POOS exercise* The data on univariate time series $y_t$ are sampled at equidistant times $t = 1, \ldots, T$. We forecast the average of $m$ future values $\bar{y}_{+1:m} = m^{-1} \sum_{t=1}^{m} y_{T+t}$. We design our POOS comparison using the following three time series (plots of the series are given in "Appendix A"),

(spret) S&P 500 value weighted daily returns including dividends available from January 2, 1926, till December 31, 2014, with a total of 23, 535 observations,

(spret2) squared daily returns, with the same period and
(tb3m) nominal interest rates for 3-month US Treasury Bills available from April 1, 1954, till August 13, 2015, with a total of 15,396 observations.

The sample size for post-WWII quarterly macroeconomic time series is $4 \times 68 = 272$ observations. We mimic the macroeconomic forecasting setup in that we use a rolling sample estimation with sample size $T = 260$ days (i.e., 1 year of daily data), and forecasting horizon $m = 20, 30, 40, 60, 90$ and $130$ days. The rolling samples are overlapping in the last (resp. first) $T - m$ observations, so that, e.g., for $m = 130$, the consecutive samples share half the observations. Hence, for the returns time series and for $m = 130$ (resp. $m = 20$), we get $N_{\text{trials}} = 178$ (resp. 1163) non-overlapping in-or-out POOS trials. All models are selected and parameters estimated anew at each forecast origin.

The simulation results in Sect. 2.2.2 have shown that *zxw* PIs have decent coverage for short-memory $e_t \sim I(0)$, but lose the coverage rapidly if the process has long memory. As a remedy, we apply an appropriate transformation before we use re-sampling and perform the reversed transformation immediately before the estimation of the kernel quantiles (see "Appendix B"). The re-sampling scheme also benefits from the prior transformation, since the stationary bootstrap is suitable for weakly dependent series. For *zxw*, we assume *spret*$\sim I(0)$, *spret2*$\sim I(d)$ with $d = 0.5$ (see Andersen et al. 2003) and *tb3m*$\sim I(1)$, and we replace $e_t$ by respective differences $de_t = (1 - L)^d e_t$ (with $L$ as lag operator). Concerning the bootstrap/analytic PIs for *arfima(p,d,q)–garch(P,Q)*, we report only the best empirical coverage probability $\hat{p}$ and corresponding relative width $\hat{w}$ among two choices of $(P, Q) \in \{(0, 0), (1, 1)\}$.

*POOS results* Similarly as in Sect. 2.2.2, we evaluate the coverage probability (2.20) and relative median width (2.21), for nominal coverage probabilities 90% (see Table 2A) and 67% (see Table 2B). Overall, the results show tight competition between *mw* and *zxw*. Better coverage probability is generally compensated by a larger width, hence less precision. Only for *tb3m*, *zxw* performs better in both aspects. The *prr* PIs show mixed performance, and it is difficult to draw any general conclusion whether one should prefer averaging of series (*series*) or averaging the forecasts (*4cast*) and whether to use analytic formulas (*anlt*) rather than bootstrapping (*boot*) to obtain PIs. We keep our focus on the coverage yield for the longer horizons.

*spret* Based on the simulation results for short-memory and heavy-tailed series, we expect that both methods of *zxw* should give decent coverage probability close to the nominal level. The real-data performance is better than suggested using artificial data, with an average drop of 9 resp. 12pp for *kernel-boot* resp. *clt-tdist* below the nominal level. On the other hand, *mw* exceed the nominal coverage even with the *naive* method. The difference in coverage probability between *robust* and *kernel-boot* reaches 15pp. The *zxw* provide advantage regarding the width, as the *robust* has twice the width of *kernel-boot* for $m = 130$. The *prr* gives decent coverage only for short horizon. For medium and long horizons, both the coverage and the width of *prr* exhibit a rapid decay. Averaging series dominates over averaging forecasts by 10pp. To our surprise, the bootstrap PIs do not outperform the analytic PIs. Regarding the width, the *mw* are by 40pp more conservative than the empirical inter-quantile range of the out-of-

**Table 2** Real-data results: comparison of coverage probabilities for *zxw*, *mw* and *prr* on each of the three daily time series: *spret*, *spret2*, *tb3m*

| | Horizon (days) | Coverage probability $\hat{p}$ | | | | | | Relative median width $\hat{w}$ | | | | | |
|---|---|---|---|---|---|---|---|---|---|---|---|---|---|
| | | 20 | 30 | 40 | 60 | 90 | 130 | 20 | 30 | 40 | 60 | 90 | 130 |
| *(A) Results of POOS forecasting experiment for nominal coverage probability $p = 1 - \alpha = 90\%$* | | | | | | | | | | | | | |
| S&P 500 returns | kernel-boot | 89.94 | 89.16 | 88.30 | 88.89 | 87.60 | 81.56 | 0.87 | 0.94 | 0.91 | 0.89 | 0.87 | 0.77 |
| | clt-tdist | 87.70 | 87.35 | 85.89 | 85.53 | 83.72 | 78.21 | 0.81 | 0.85 | 0.84 | 0.81 | 0.79 | 0.70 |
| | robust | 92.00 | 94.58 | 94.32 | 94.83 | 96.90 | 96.09 | 1.06 | 1.37 | 1.22 | 1.46 | 1.57 | 1.43 |
| | bayes | 87.96 | 92.39 | 90.71 | 93.28 | 92.64 | 93.30 | 0.92 | 1.19 | 1.02 | 1.21 | 1.29 | 1.19 |
| | naive | 86.07 | 90.19 | 88.47 | 92.25 | 92.25 | 84.92 | 0.83 | 1.07 | 0.90 | 1.06 | 1.09 | 0.97 |
| | series-anlt | 85.21 | 84.52 | 84.51 | 81.14 | 79.84 | 75.98 | 0.79 | 0.84 | 0.83 | 0.80 | 0.78 | 0.69 |
| | series-boot | 85.12 | 85.16 | 83.30 | 81.40 | 79.84 | 73.74 | 0.77 | 0.83 | 0.82 | 0.78 | 0.77 | 0.68 |
| | 4cast-anlt | 83.32 | 80.39 | 81.41 | 77.78 | 68.22 | 68.72 | 0.88 | 0.90 | 0.85 | 0.77 | 0.69 | 0.57 |
| | 4cast-boot | 82.46 | 80.26 | 81.76 | 72.09 | 67.44 | 62.57 | 0.86 | 0.88 | 0.84 | 0.77 | 0.70 | 0.57 |
| S&P 500 returns² | kernel-boot | 92.61 | 91.87 | 91.05 | 90.44 | 90.31 | 87.15 | 0.39 | 0.37 | 0.36 | 0.39 | 0.36 | 0.31 |
| | clt-tdist | 91.49 | 93.03 | 92.08 | 91.99 | 91.86 | 89.39 | 0.43 | 0.43 | 0.41 | 0.46 | 0.43 | 0.37 |
| | robust | 89.51 | 89.42 | 87.78 | 87.60 | 87.21 | 86.03 | 0.35 | 0.34 | 0.30 | 0.33 | 0.29 | 0.26 |
| | bayes | 88.39 | 88.26 | 85.20 | 86.30 | 84.50 | 83.24 | 0.32 | 0.32 | 0.27 | 0.31 | 0.28 | 0.24 |
| | naive | 87.79 | 88.77 | 81.76 | 83.46 | 78.29 | 68.16 | 0.30 | 0.30 | 0.21 | 0.24 | 0.19 | 0.13 |
| | series-anlt | 41.44 | 33.55 | 28.92 | 23.77 | 16.67 | 13.97 | 0.10 | 0.08 | 0.07 | 0.07 | 0.06 | 0.04 |
| | series-boot | 75.00 | 75.00 | 70.89 | 67.00 | 62.69 | 52.27 | 0.32 | 0.27 | 0.21 | 0.31 | 0.25 | 0.06 |
| | 4cast-anlt | 72.74 | 70.45 | 66.95 | 59.59 | 57.75 | 44.13 | 0.19 | 0.18 | 0.14 | 0.14 | 0.12 | 0.09 |
| | 4cast-boot | 78.66 | 77.18 | 70.78 | 65.08 | 58.67 | 51.67 | 0.23 | 0.17 | 0.24 | 0.13 | 0.23 | 0.10 |

Table 2 continued

| | | Coverage probability $\hat{p}$ | | | | | | Relative median width $\hat{w}$ | | | | | |
|---|---|---|---|---|---|---|---|---|---|---|---|---|---|
| | Horizon (days) | 20 | 30 | 40 | 60 | 90 | 130 | 20 | 30 | 40 | 60 | 90 | 130 |
| TB3M interest rate | kernel-boot | 89.29 | 88.69 | 85.45 | 85.32 | 82.74 | 75.86 | 0.05 | 0.06 | 0.07 | 0.08 | 0.11 | 0.11 |
| | clt-tdist | 92.33 | 92.46 | 89.95 | 89.68 | 86.31 | 81.90 | 0.06 | 0.07 | 0.08 | 0.09 | 0.12 | 0.13 |
| | robust | 90.34 | 85.91 | 86.24 | 80.16 | 73.21 | 72.41 | 0.08 | 0.08 | 0.09 | 0.09 | 0.11 | 0.12 |
| | bayes | 90.34 | 85.52 | 85.45 | 79.37 | 73.21 | 72.41 | 0.08 | 0.08 | 0.09 | 0.09 | 0.11 | 0.12 |
| | naive | 64.42 | 61.11 | 45.50 | 42.86 | 32.14 | 25.00 | 0.15 | 0.15 | 0.11 | 0.11 | 0.09 | 0.07 |
| | series-anlt | 84.13 | 81.94 | 79.10 | 75.40 | 76.19 | 68.10 | 0.05 | 0.06 | 0.07 | 0.08 | 0.11 | 0.13 |
| | series-boot | 63.93 | 68.38 | 60.49 | 64.41 | 57.89 | 47.06 | 0.06 | 0.06 | 0.08 | 0.08 | 0.08 | 0.10 |
| | 4cast-anlt | 83.07 | 80.95 | 71.69 | 61.11 | 52.98 | 36.21 | 0.05 | 0.06 | 0.07 | 0.07 | 0.09 | 0.09 |
| | 4cast-boot | 42.46 | 41.67 | 41.80 | 47.22 | 45.83 | 45.69 | 0.08 | 0.10 | 0.10 | 0.13 | 0.15 | 0.16 |

*(B) Results of POOS forecasting experiment for nominal coverage probability $p = 1 - \alpha = 67\%$*

| | | | | | | | | | | | | | |
|---|---|---|---|---|---|---|---|---|---|---|---|---|---|
| S&P 500 returns | kernel-boot | 67.50 | 68.52 | 67.30 | 66.67 | 63.95 | 61.45 | 1.01 | 0.97 | 1.01 | 1.00 | 0.96 | 0.87 |
| | clt-tdist | 65.43 | 65.03 | 64.37 | 61.50 | 58.91 | 60.34 | 0.94 | 0.89 | 0.92 | 0.92 | 0.88 | 0.80 |
| | robust | 71.28 | 79.61 | 80.21 | 84.75 | 84.11 | 82.12 | 1.16 | 1.34 | 1.37 | 1.70 | 1.83 | 1.69 |
| | bayes | 66.38 | 74.32 | 68.50 | 73.13 | 75.19 | 72.63 | 1.03 | 1.18 | 1.05 | 1.27 | 1.28 | 1.17 |
| | naive | 63.89 | 72.65 | 64.20 | 70.03 | 70.93 | 68.16 | 0.93 | 1.07 | 0.95 | 1.16 | 1.17 | 1.08 |
| | series-anlt | 60.19 | 59.74 | 58.69 | 56.59 | 52.33 | 58.10 | 0.82 | 0.79 | 0.83 | 0.82 | 0.80 | 0.73 |
| | series-boot | 62.51 | 61.03 | 61.45 | 57.11 | 53.49 | 57.54 | 0.88 | 0.83 | 0.88 | 0.87 | 0.85 | 0.77 |
| | 4cast-anlt | 56.92 | 55.74 | 56.80 | 53.49 | 49.61 | 45.25 | 0.96 | 0.87 | 0.87 | 0.83 | 0.74 | 0.61 |
| | 4cast-boot | 59.07 | 55.61 | 56.28 | 51.16 | 46.90 | 38.55 | 0.99 | 0.91 | 0.90 | 0.86 | 0.77 | 0.65 |

**Table 2** continued

| | Horizon (days) | Coverage probability $\hat{p}$ | | | | | | Relative median width $\hat{w}$ | | | | | |
|---|---|---|---|---|---|---|---|---|---|---|---|---|---|
| | | 20 | 30 | 40 | 60 | 90 | 130 | 20 | 30 | 40 | 60 | 90 | 130 |
| S&P 500 returns² | kernel-boot | 76.61 | 78.45 | 78.31 | 79.33 | 78.68 | 77.65 | 0.60 | 0.64 | 0.63 | 0.63 | 0.59 | 0.63 |
| | clt-tdist | 81.77 | 81.29 | 82.44 | 82.43 | 82.17 | 79.89 | 0.78 | 0.80 | 0.78 | 0.77 | 0.71 | 0.72 |
| | robust | 75.84 | 77.03 | 74.35 | 74.94 | 71.32 | 66.48 | 0.61 | 0.65 | 0.58 | 0.59 | 0.52 | 0.49 |
| | bayes | 73.95 | 74.58 | 70.91 | 70.28 | 67.44 | 62.57 | 0.58 | 0.60 | 0.51 | 0.52 | 0.44 | 0.42 |
| | naive | 71.97 | 73.42 | 57.31 | 60.98 | 51.94 | 44.69 | 0.55 | 0.57 | 0.41 | 0.41 | 0.32 | 0.27 |
| | series-anlt | 21.32 | 14.97 | 12.74 | 8.79 | 6.59 | 6.15 | 0.14 | 0.12 | 0.11 | 0.09 | 0.07 | 0.07 |
| | series-boot | 50.31 | 46.82 | 48.73 | 45.00 | 32.84 | 31.82 | 0.36 | 0.35 | 0.31 | 0.38 | 0.26 | 0.13 |
| | 4cast-anlt | 45.06 | 41.55 | 39.41 | 36.01 | 33.72 | 25.14 | 0.28 | 0.26 | 0.21 | 0.18 | 0.16 | 0.13 |
| | 4cast-boot | 58.84 | 55.19 | 51.95 | 42.06 | 36.00 | 31.67 | 0.50 | 0.39 | 0.36 | 0.36 | 0.25 | 0.29 |
| TB3M interest rate | kernel-boot | 69.18 | 67.86 | 66.93 | 61.11 | 57.14 | 59.48 | 0.05 | 0.06 | 0.07 | 0.08 | 0.10 | 0.12 |
| | clt-tdist | 76.32 | 74.40 | 71.16 | 67.86 | 63.10 | 62.93 | 0.06 | 0.07 | 0.08 | 0.09 | 0.11 | 0.14 |
| | robust | 71.83 | 65.87 | 67.46 | 56.75 | 50.00 | 56.03 | 0.08 | 0.08 | 0.09 | 0.09 | 0.11 | 0.12 |
| | bayes | 71.56 | 65.67 | 66.93 | 56.75 | 50.00 | 56.03 | 0.08 | 0.08 | 0.09 | 0.09 | 0.11 | 0.12 |
| | naive | 34.39 | 32.54 | 22.49 | 18.65 | 18.45 | 13.79 | 0.14 | 0.14 | 0.11 | 0.10 | 0.08 | 0.07 |
| | series-anlt | 57.67 | 59.13 | 55.03 | 55.16 | 51.19 | 51.72 | 0.04 | 0.05 | 0.06 | 0.07 | 0.09 | 0.11 |
| | series-boot | 48.09 | 50.43 | 49.38 | 49.15 | 28.95 | 29.41 | 0.05 | 0.06 | 0.06 | 0.09 | 0.10 | 0.09 |
| | 4cast-anlt | 55.82 | 56.55 | 49.47 | 42.06 | 28.57 | 20.69 | 0.04 | 0.05 | 0.06 | 0.07 | 0.08 | 0.09 |
| | 4cast-boot | 37.17 | 35.12 | 34.66 | 40.48 | 35.12 | 36.21 | 0.08 | 0.10 | 0.10 | 0.13 | 0.15 | 0.17 |

The reported values are coverage probabilities and median widths of the respective prediction intervals based on 100+ rolling pseudo-out-of-sample trials. The median widths are reported relative to the width of the respective inter-quantile range computed from the empirical distribution of the targeted out-of-sample averages sample mean-returns, whereas *zxw* resp. *prr* are 23–30pp resp. 31–43pp below the inter-quantile range width.

*spret2* Realized volatility is known for the persistence and heavy tails. The *mw* give slightly lower coverage probabilities than *zxw* compensated by a relatively smaller width, thus better precision. With the growing horizon all *prr* methods suffer a drop in coverage, at least 40pp below the nominal level, accompanied by the largest reduction of width among all methods. The bootstrap PIs dominate over analytic, and the competition between *4cast* and *series* is tight. Concerning the relative width, when compared to the previous case of returns, all methods provide very narrow PIs. We believe that the seemingly shrinking width of PIs is caused by a larger dispersion of the entire *spret2* [entering the denominator of (2.21)] compared to the dispersion of each local average [the nominator in (2.21)]. Note that for *spret2* the denominator in (2.21) does not provide adequate scale and the discrepancy will become even worse for more persistent *tb3m*.

*tb3m* Interest rates exhibit strong persistence, and for enhanced performance of *zxw*, we again apply differencing, but with $d = 1$. Note that the *naive* PIs have coverage probabilities as low as 25%. The coverage probabilities of all methods are lower than for the last two series, but *zxw* performs better than *mw* for all horizons. Moreover, *zxw* gives better results in terms of width. The coverage probability for *prr* falls far below the nominal level as the horizon grows. Here, *series* dominates *4cast*, and bootstrap PIs are inferior to the analytic PIs, even though with only half the nominal coverage.

Except *spret*, *clt-tdist* gives slightly higher coverage probabilities than *kernel-boot* corresponding to smaller precision. Eventually, we prefer the *kernel-boot* method and use it in the following section for computing PIs for eight economic time series and S&P 500 returns.

## 4 Prediction intervals for economic series' growth rates and S&P 500 returns

Müller and Watson (2016), in Table 5 on pages 1731–1732, gave their long-run PIs for eight quarterly post-WWII US economic time series and quarterly returns. Since our *kernel-boot* method performed well in the last real-data POOS comparison, we employ this method in order to obtain alternative PIs for these series. We report these PIs in Tables 3 and 4. Müller and Watson (2016) compare their PIs to those published by CBO. They conclude that some similarities between their PIs for series such as GDP are due to a combination of (i) CBO's ignorance for parameter uncertainty and (ii) CBO's ignorance of possible anti-persistence of GDP during Great Moderation. Since the effects of (i) and (ii) on the PIs width are the opposite, they eventually seem to cancel out.

The eight economic time series are real per capita GDP, real per capita consumption expenditures, total factor productivity, labor productivity, population, inflation (PCE[7]), inflation (CPI[8]) and Japanese inflation (CPI)—all transformed into log-differences.

---

[7] Personal consumption expenditure deflator.

[8] Consumer price index.

**Table 3** Prediction intervals for long-run averages of quarterly post-WWII growth rates

| Horizon (years) | $\hat{p} = 67\%$ | | | $\hat{p} = 90\%$ | | |
|---|---|---|---|---|---|---|
| | 10 | 25 | 50 | 10 | 25 | 50 |
| GDP/Pop | [−0.88, 4.65] | [−1.11, 4.76] | [−0.87, 4.68] | [−2.97, 6.78] | [−2.96, 6.59] | [−2.86, 6.48] |
| Cons/Pop | [0.56, 3.45] | [0.60, 3.45] | [0.59, 3.49] | [−0.54, 4.53] | [−0.43, 4.38] | [−0.40, 4.55] |
| TF prod. | [−0.46, 2.92] | [−0.37, 2.96] | [−0.40, 2.76] | [−1.61, 4.16] | [−1.55, 4.02] | [−1.49, 3.83] |
| Labor prod. | [0.89, 3.42] | [0.84, 3.24] | [0.90, 3.37] | [−0.11, 4.35] | [0.06, 4.11] | [0.08, 4.15] |
| Population | [0.44, 0.95] | [0.25, 1.00] | [−0.11, 0.90] | [0.24, 1.17] | [−0.06, 1.35] | [−0.50, 1.35] |
| PCE infl. | [−4.06, 2.32] | [−6.01, 3.83] | [−9.50, 5.15] | [−7.39, 4.70] | [−9.74, 7.40] | [−14.38, 9.86] |
| CPI infl. | [−4.75, 1.61] | [−6.32, 2.04] | [−9.43, 3.29] | [−9.00, 4.03] | [−10.54, 6.57] | [−14.95, 7.46] |
| Jap. CPI infl. | [−5.20, 2.79] | [−7.12, 4.18] | [−8.72, 5.85] | [−8.10, 7.63] | [−11.38, 10.07] | [−14.51, 12.19] |
| Returns | [2.20, 12.20] | [3.50, 10.75] | [4.78, 10.22] | [−1.93, 15.69] | [0.88, 12.95] | [2.90, 11.71] |

The eight macroeconomic time series are *real per capita GDP*, *real per capita consumption expenditures*, *total factor productivity*, *labor productivity*, *population*, *personal consumption expenditure deflator*, *CPI inflation* and *Japanese inflation*—all transformed into log-differences. The prediction intervals provide alternative to Table 5 of Müller and Watson (2016), who report intervals also for the horizon $m = 75$ years for these short post-WWII quarterly series. However, since this horizon would exceed the sample size, we cannot provide the *kernel-boot* as alternative. But we present results for this horizon in the next table, where we use longer yearly series

**Table 4** Prediction intervals for long-run averages of annual growth rates and annual S&P 500 returns

| Horizon (years) | | 10 | 25 | 50 | 75 |
| --- | --- | --- | --- | --- | --- |
| 67% | GDP/Pop | [−1.43, 5.61] | [−1.59, 5.68] | [−1.85, 5.65] | [−1.72, 5.36] |
| | Cons/Pop | [−1.07, 4.27] | [−1.15, 4.41] | [−0.96, 4.33] | [−1.08, 4.26] |
| | Population | [0.33, 0.99] | [0.08, 1.11] | [−0.21, 1.16] | [−0.54, 1.15] |
| | CPI infl. | [−2.72, 6.02] | [−2.80, 6.21] | [−3.19, 6.69] | [−5.27, 9.46] |
| | Returns | [0.38, 13.61] | [3.74, 10.68] | [3.60, 9.67] | [4.44, 8.29] |
| 90% | GDP/Pop | [−5.00, 8.44] | [−4.30, 8.47] | [−4.92, 8.24] | [−4.49, 7.96] |
| | Cons/Pop | [−3.12, 6.21] | [−3.03, 6.22] | [−2.80, 6.03] | [−2.90, 6.27] |
| | Population | [0.13, 1.23] | [−0.24, 1.51] | [−0.63, 1.74] | [−1.13, 1.81] |
| | CPI infl. | [−6.02, 12.65] | [−9.00, 12.13] | [−8.13, 12.87] | [−11.26, 16.13] |
| | Returns | [−3.64, 17.50] | [0.45, 12.62] | [1.61, 11.77] | [2.82, 9.49] |

The macroeconomic time series are *real per capita GDP*, *real per capita consumption expenditures*, *population*, *CPI inflation* and *Japanese inflation*—all transformed into log-differences. This table provides alternative prediction intervals to those reported in Table 5 of Müller and Watson (2016)

(Plots are given in "Appendix A.") The data are available from 1Q-1947 till 4Q-2014, and we forecast them over next $m = 10, 25$ and $50$ years. For a subset of these series, we report results based on longer (yearly) sample starting in 1Q-1920, and we add the horizon $m = 75$ years for these yearly series.

For *per capita real GDP*, *per capita consumption* and *productivity*, we use differencing with $d = 0.5$ for the *kernel-boot* PIs. Thus, these intervals are wider than in Müller and Watson (2016), especially those for GDP. This case is similar to the case of realized volatility in the previous section. Wide PIs are often considered as a failure of the forecasting method or model. On the other hand, they can also reflect the higher uncertainty about the series future. The width of PIs for GDP is not surprising given that similar as CBO, we do not account for the possible anti-persistence during the Great Moderation. With the longer yearly sample, our PIs get even wider, as a result of higher volatility in the early twentieth century. Interestingly, the growth in Labor production seems to be higher in general than reported by Müller and Watson (2016).

*Consumption, population and inflation* are well known as quite persistent. Therefore, we would expect that similarly as in case of interest rates, *kernel-boot* could give better coverage and possibly narrower PIs than *robust*. The uncertainty is similarly large according to both our *kernel-boot* and *robust*, but the location is generally shifted downward, especially for inflation, where the shift is about −2pp compared to Müller and Watson (2016).

Finally, for the *quarterly returns*, we might expect *kernel-boot* to give less conservative thus narrow estimates, and we see this happening with discrepancy growing with the forecasting horizons. It is clear that *robust* is very conservative in uncertainty about positive returns, where the difference from *kernel-boot* reached 11pp. Employing the longer yearly time series makes the difference fall to 3pp. On the other hand, 3pp is a lot from an investors perspective.

## 5 Discussion

We have constructed prediction intervals for univariate economic time series. Our forecasting comparison shows that even the simple methods of Zhou et al. (2010) provide a valid alternative for sophisticated prediction intervals designed specifically for the economic framework by Müller and Watson (2016). However, based on our simulation results, we emphasize that both the methods and the series need to be suitably adjusted, especially under the small sample constraint, which, on the other hand, is quite common in practice. Based on the comparison results, we eventually provided alternative long-run prediction intervals for eight US economic indicators.

Forecasting average growth of economic series over the coming decades is a very ambitious task, and naturally, there are doubts about its usefulness in practice. The test of Breitung and Knüppel (2018), whether a forecast is informative, is based on the prediction error variance. They conclude that economic forecasts beyond a horizon of several quarters become uninformative. At first sight, such a claim seems to be an argument against following the research of Müller and Watson (2016) and Zhou et al. (2010). However, there are some differences in the assumptions and targets which have to be carefully analyzed before we make such statements. The assumption of a long-memory component is crucial, and it is hard to verify and distinguish it from a possible structural break. In our paper, we did not tackle the issue of whether long-term predictions are informative or not. We instead probed into the existing methods and provided new empirical comparison results.

Throughout this paper, we focused on PIs estimated from historical data on the predicted series. A multivariate or high-dimensional extension would, of course, be attractive. It is widely recognized that big data contain additional forecasting power. Unfortunately, in the economic literature, the boom of forecasting with many predictors (e.g., Stock and Watson 2012; Elliott et al. 2013; Kim and Swanson 2014) is mainly focused on short horizons and point-forecasting (for an exception see Bai and Ng 2006). This is not a coincidence. Many economic time series exhibit persistence (of varying degrees), and this is their essential property in the long run. These long-term effects, combined over many series, are difficult to understand, partially due to their dependence on unknown nuisance parameters (see Elliott et al. 2015). The role of cointegration in long-run forecasting is investigated by Christoffersen and Diebold (1998).

We do not use some methods such as quantile (auto-) regression (Koenker 2005) in the current study, and the out-of-sample forecasting comparison could be enhanced by statistical tests (see Clements and Taylor 2003; Gneiting and Raftery 2007, for example).

An extension (including the theory) of Zhou et al. (2010) into a high-dimensional regression framework using the LASSO estimator is currently being developed. Even more challenging is a case of multivariate target series and subsequent construction of simultaneous prediction intervals which can have interesting implications for market trading strategies.

**Acknowledgements** Open access funding provided by University of Vienna. We would like to thank Jörg Breitung, Francis Diebold, Alexander Kment, Lubos Pastor, Irina Pimenova, Justin Veenstra, Mark Watson as well as the participants of the conference: Big Data in Predictive Dynamic Econometric Modeling held at the University of Pennsylvania and the 1st Vienna Workshop on Economic Forecasting 2018 held at the Institute for Advanced Studies for helpful discussion and for answering our questions. Special thanks go to Eric Nesbitt and the two anonymous referees. We also acknowledge the computational resources provided by the Vienna Scientific Cluster. Note that the opinions expressed in the paper are those of the authors and do not necessarily reflect the opinions of the Institute for Financial Policy.

**Compliance with ethical standards**

**Conflict of interest** The authors declare that they have no conflict of interest.

**Funding** The research was partly supported by grant NSF/DMS 1405410. M. Chudý received financial support from J.W. Fulbright Commission for Educational Exchange in the Slovak Republic, The Ministry of Education, Science, Research and Sport of the Slovak Republic and the Stevanovich Center for Financial Mathematics.

## Appendix A: Figures of time series used in Sects. 3 and 4

See Figs. 1, 2 and 3

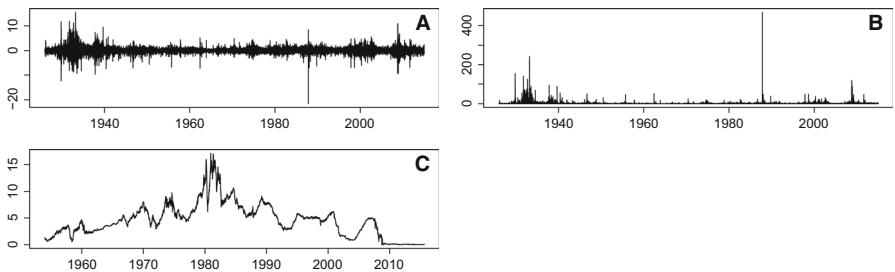

**Fig. 1** Daily time series: **a** S&P 500 value weighted daily returns incl. dividend, **b** squared returns, **c** nominal interest rates for 3-month US Treasury Bills

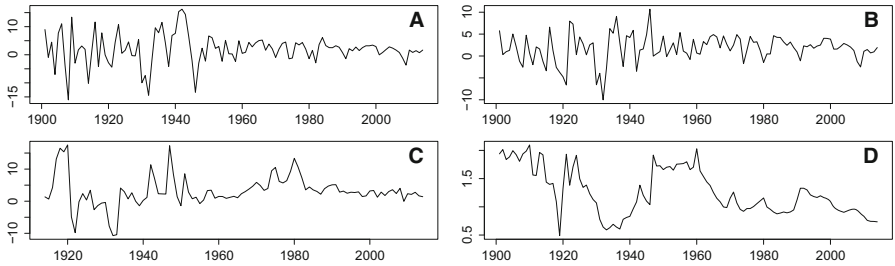

**Fig. 2** Annual time series—growth rates: **a** real per capita GDP, **b** real per capita consumption expenditures, **c** inflation (CPI), **d** population

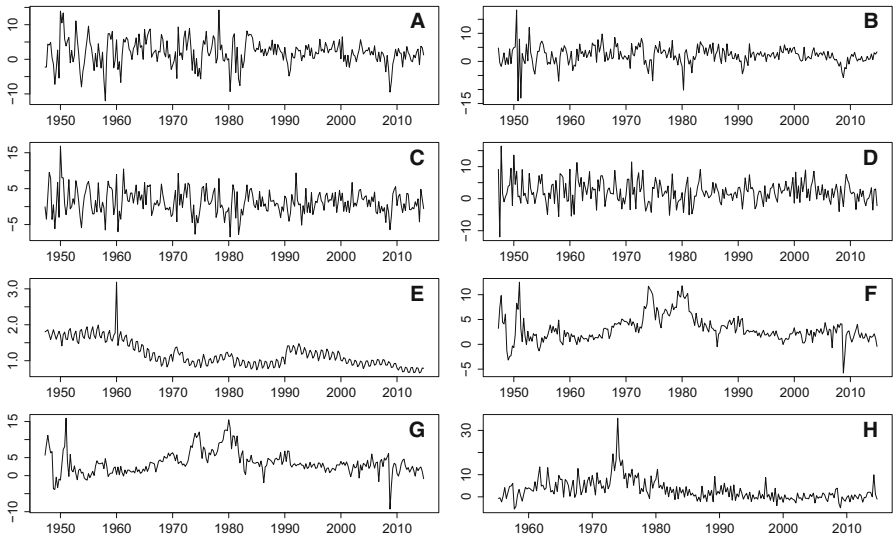

**Fig. 3** Quarterly time series—growth rates: **a** real per capita GDP, **b** real per capita consumption expenditures, **c** total factor productivity, **d** labor productivity, **e** population, **f** prices (PCE), **g** inflation (CPI), **h** Japanese inflation

## Appendix B: Additional steps for implementation of zxw and mw

All macro-series in Sect. 4 are transformed to log-differences. This does not preclude long-memory dynamics or even a unit root. Note that if $y_t$ is $I(1)$ and has deterministic trend component rather than a constant level, the location of the PI would have to be shifted to $\frac{m+1}{2}\overline{\Delta y}$ instead of $\bar{y}$.

*kernel-boot:*

1. Compute the mean adjusted series $e_t = y_t - \bar{y}$, $t = 1, \ldots, T$.
2. Fix $d = 0.5$ or $d = 1$ and compute the difference series $de_t = (1 - L)^d$, $t = 2, \ldots, T$, where $L$ denotes lag operator.
3. Replicate $de_t$, $t = 2, \ldots, T$ $B$ times getting $de_t^b$, $t = 2, \ldots, T$, $b = 1, \ldots, B$.

4. Compute the series of overlapping rolling means $\bar{de}^b_{t(m)} = m^{-1}\sum_{i=1}^{m} de^b_{t-i+1}$, $t = m, \ldots T$ from every replicated series.
5. Estimate quantiles $\hat{Q}(\alpha/2)$ and $\hat{Q}(1-\alpha/2)$ from $\bar{e}^b_{T(m)}, b = 1, \ldots, B$ with $T = 260$ obtained as $\bar{e}^b_{T(m)} = m^{-1}\sum_{i=1}^{m}(1-L)^{-d} de^b_{T-i+1}$.
6. The PI is given by $[L, U] = \bar{y} + [Q^t_{\kappa-1}(\alpha/2), Q^t_{\kappa-1}(1-\alpha/2)]\sigma/\sqrt{m}$.

*clt-tdist:*

1. Compute the mean adjusted series $e_t = y_t - \bar{y}, t = 1, \ldots, T$.
2. Fix $d = 0.5$ or $d = 1$ and compute the difference series $de_t = (1-L)^d, t = 2, \ldots, T$, where $L$ denotes lag operator.
3. Estimate the long-run standard deviation $\tilde{\sigma}$ of $de_t, t = 2, \ldots, T$.
4. Compute the long-run standard deviation of $e_t$: for $d = 1$, $\sigma_e(\tilde{\sigma}) = \tilde{\sigma}\sqrt{(m+1)/2}$ and for[9] $d = 0.5$, $\sigma_e(\tilde{\sigma}) = \tilde{\sigma} m^{-1}\sqrt{\sum_{i=1}^{m}(\sum_{j=0}^{m-i}(-1)^j \binom{-0.5}{j})^2}$.
5. The PI is given by $\bar{y} + [Q^t_{\kappa-1}(\alpha/2), Q^t_{\kappa-1}(1-\alpha/2)]\sigma_e$.

*robust* after steps 1–4.

5.1 Compute weights for specific choice of $q$ and $m/T$ and the prior from step 3.
5.2 Numerically approximate s. c. least favorable distribution (LFD) of $\theta$ for specific choice of $q$ and $m/T$ (see "Appendix" of Müller and Watson (2016)).
5.3 Using the weights and the LFD solve the minimization problem (14) on page 1721 in Müller and Watson (2016) to get quantiles which give uniform coverage and minimize the expected PIs width.
6. The same as in *bayes* with the robust quantiles.

## Appendix C: Derivation of an error standard deviation for time-aggregated forecast

The formula for computing prediction error sd is $\hat{as}_{T,+1:m} = \frac{1}{m}\sqrt{\sum_{i=1}^{m}(\hat{\sigma}_{T,T+i}\sum_{j=0}^{m-i}\hat{\Psi}_j)^2}$, where $\hat{\Psi}_0 = 1$ and $\hat{\Psi}_j, j = 2, \ldots, m$ are the estimates of coefficients from the causal representation of $y_t$. The $\hat{\sigma}_{T,T+i}$ is the *garch* forecast for innovations deviation. For simplicity, we show the derivation for the case of constant innovation variance.

Assume that $y_t$ has causal representation:

$$y_t = \epsilon_t + \Psi_1 \epsilon_{t-1} + \cdots,$$

where $\epsilon_t \sim (0, \sigma^2)$ is the innovation process with constant second moment. Standing at time $T$ the $i$th-step-ahead prediction error can be expressed as

$$pe_{T,i} = \epsilon_{t+i} + \Psi_1 \epsilon_{t+i-1} + \cdots + \Psi_{i-1}\epsilon_{T+1}.$$

---

[9] See http://mathworld.wolfram.com/BinomialSeries.html.

The average prediction error over horizons $i = 1, \ldots, m$ is therefore given by

$$\bar{p}e_{T,+1:m} = \frac{1}{m} \sum_{i=1}^{m} \sum_{j=i}^{1} \Psi_{i-j} \epsilon_{T+j},$$

with $\Psi_0 = 1$. Now, this can be rewritten as

$$\bar{p}e_{T,+1:m} = \frac{1}{m} \left( \epsilon_{T+1} \underbrace{\sum_{j=0}^{m-1} \Psi_j}_{c_{m-1}} + \epsilon_{T+2} \underbrace{\sum_{j=0}^{m-2} \Psi_j}_{c_{m-2}} + \cdots + \epsilon_{T+m-1} \underbrace{\sum_{j=0}^{1} \Psi_j}_{c_1} + \epsilon_{T+m} \right),$$

where $c_0 = \Psi_0 = 1$. Since innovations are uncorrelated, we can compute the variance of average prediction error over the horizons $i = 1, \ldots, m$ as

$$\text{var}(\bar{p}e_{T,+1:m}) = \left(\frac{\sigma}{m}\right)^2 \sum_{i=1}^{m} c_{m-i}^2.$$

## Appendix D: Discussion on CLT and QTL methods

For the interested reader here we provide some discussion on the justification of the two original methods from Zhou et al. (2010) and how one can verify them in linear and possibly nonlinear processes. First, we discuss a result for the CLT method. Assume the process $e_t$ admits the following linear form

$$e_t = \sum_{j=0}^{\infty} a_j \epsilon_{t-j}, \tag{5.1}$$

where $\epsilon_t$ are mean-zero, independent and identically distributed (i.i.d.) random variables with finite second moment. For this structural form, we can evaluate (2.14). We assume a particular decay rate of $a_i$ and state the following theorem.

**Theorem 1** *Assume the process $e_t$ admits representation* (5.1) *where $a_i$ satisfies*

$$a_i = O(i^{-\chi} (\log i)^{-A}), \quad \chi > 1, A > 0, \tag{5.2}$$

*where larger $\chi$ and $A$ mean fast decay rate of dependence. Further assume, $A > 5/2$ if $1 < \chi < 3/2$. Then the sufficient condition* (2.14) *implies that convergence* (2.15) *to the asymptotic normal distribution holds.*

*Proof*

$$\|\mathbb{E}(S_m|\mathcal{F}_0)\|^2 = \|(a_1 + \cdots + a_m)\epsilon_0 + (a_2 + \cdots + a_m)\epsilon_{-1} + \cdots\|^2 = \sum_{i=1}^{m} b_i^2, \quad (5.3)$$

where $b_i = a_i + \cdots + a_m$. Note that $\sum_{i=1}^{m} b_i^2$ assumes the following value depending on $\chi > 3/2$ or not. Thus, (2.14) holds since by elementary calculations,

$$\sum_{i=1}^{m} b_i^2 = \begin{cases} O(m^{3-2\chi}(\log m)^{-2A}), & \text{for } 3 - 2\chi > 0 \\ O(1) & \text{for } 3 - 2\chi \leq 0. \end{cases} \quad (5.4)$$

□

Note that Theorem 1 concerns only linear processes. This class covers a large class of time series processes already. However, we do not necessarily require linearity of the error process $(e_t)$. One can equivalently use the functional dependence measure introduced in Wu (2005) to state an equivalent result for stationary possibly nonlinear error processes of the form

$$e_t = G(\epsilon_t, \epsilon_{t-1}, \ldots),$$

where $\epsilon_i$ are i.i.d. random variables. For this process assuming $p \geq 2$ moments one can define the functional dependence measure

$$\delta_{j,p} = \|e_j - G(\epsilon_j, \ldots, \epsilon_0^*, \ldots)\|_p,$$

where $\epsilon_{(\cdot)}^*$ is an i.i.d. copy of $\epsilon_{(\cdot)}$ process. For the specific case of $e_t$ assuming a linear form as specified in (5.1), we have $\delta_{j,p} = a_j$. This lays down a straightforward way in how our results for the linear process can be easily extended to nonlinear processes.

Next, for the sake of completeness, we borrow a result from Zhou et al. (2010) that discusses the quantile consistency for the QTL method. Recall that we will exhibit as promised that this method allows for the situation where the i.i.d. innovations $\epsilon_t$ in the decomposition (5.1) can have both light tails, i.e., $\mathbb{E}(|\epsilon_t|^2) < \infty$, and heavy tails, i.e., $\alpha < 2$ where $\alpha = \sup_{r>0}\{r : \mathbb{E}(|\epsilon_t|^r) < \infty\}$.

We will impose the following conditions on the coefficients for short- or long-range dependence and also assume boundedness of the density of $\epsilon_t$ in the following sense:

$$\text{(SRD)} : \sum_{j=0}^{\infty} |a_j| < \infty,$$

$$\text{(DEN)} : \sup_{x \in \mathbb{R}} f_\epsilon(x) + |f_\epsilon'(x)| < \infty,$$

$$\text{(LRD)} : a_j = O((j+1)^{-\gamma} l(j)), 1/\alpha < \gamma$$
$$< 1, l(\cdot) \text{ is slowly varying function (s. v. f.)}, \quad (5.5)$$

where s. v. f. is a function $g(x)$ such that $\lim_{x \to \infty} g(tx)/g(x) = 1$ for any $t$. The condition (SRD) is a classic short-range-dependent condition (see Box et al. 2015, for more discussion). (LRD) refers to the long memory of the time series, and it is satisfied by a large class of models such as *arfima*. (DEN) is also a mild condition since by inversion theorem, all symmetric stable distributions fall under this condition. We borrow the following result from Zhou et al. (2010) for linear process. It is worth noting that one can extend this to nonlinear processes as well by defining the coupling-based dependence on predictive density of $e_t$ as done in Zhang and Wu (2015), but we postpone that discussion for a future paper.

*Quantile consistency results for the quantile method* For a fixed $0 < q < 1$, let $\hat{Q}(q)$ and $\tilde{Q}(q)$ denote the $q$th sample quantile and actual quantile of $(m \bar{e}_{t(m)}/H_m)$; $t = m, \ldots, T$ where, using (5.5),

$$H_m = \begin{cases} \sqrt{m}, & \text{if (SRD) holds and } \mathbb{E}(\epsilon_j^2) < \infty, \\ \inf \left\{ x : \mathbb{P}(|\epsilon_i| > x) \leq \frac{1}{m} \right\} & \text{if (SRD) holds and } \mathbb{E}(\epsilon_j^2) = \infty, \\ m^{3/2-\gamma} l(m) & \text{if (LRD) holds and } \mathbb{E}(\epsilon_j^2) < \infty, \\ \inf\{x : \mathbb{P}(|\epsilon_i| > x) m^{1-\gamma} l(m) & \text{if (LRD) holds and } \mathbb{E}(\epsilon_j^2) = \infty. \end{cases} \quad (5.6)$$

We have the following different rates of convergence of quantiles based on the nature of tail or dependence:

**Theorem 2** (Zhou et al. (2010) Th 1:4)[Quantile consistency result]

– *Light-tailed (SRD) Suppose (DEN) and (SRD) hold and $\mathbb{E}(\epsilon_j^2) < \infty$. If $m^3/T \to 0$, then for any fixed $0 < q < 1$,*

$$|\hat{Q}(q) - \tilde{Q}(q)| = O_{\mathbb{P}}(m/\sqrt{T}). \quad (5.7)$$

– *Light-tailed (LRD) Suppose (LRD) and (DEN) hold with $\gamma$ and $l(\cdot)$ in (5.5). If $m^{5/2-\gamma} T^{1/2-\gamma} l^2(T) \to 0$, then for any fixed $0 < q < 1$,*

$$|\hat{Q}(q) - \tilde{Q}(q)| = O_{\mathbb{P}}(m T^{1/2-\gamma} |l(T)|). \quad (5.8)$$

– *Heavy-tailed (SRD) Suppose (DEN) and (SRD) hold and $\mathbb{E}(|\epsilon_j|^\alpha) < \infty$ for some $1 < \alpha < 2$. If $m = O(T^k)$ for some $k < (\alpha - 1)/(\alpha + 1)$, then for any fixed $0 < q < 1$,*

$$|\hat{Q}(q) - \tilde{Q}(q)| = O_{\mathbb{P}}(m T^\nu) \text{ for all } \nu > 1/\alpha - 1. \quad (5.9)$$

– *Heavy-tailed (LRD) Suppose (LRD) holds with $\gamma$ and $l(\cdot)$ in (5.5). If $m = O(T^k)$ for some $k < (\alpha\gamma - 1)/(2\alpha + 1 - \alpha\gamma)$, then for any fixed $0 < q < 1$,*

$$|\hat{Q}(q) - \tilde{Q}(q)| = O_{\mathbb{P}}(m T^\nu) \text{ for all } \nu > 1/\alpha - \gamma. \quad (5.10)$$

## References

Andersen TG, Bollerslev T, Diebold FX, Labys P (2003) Modeling and forecasting realized volatility. Econometrica 71(2):579–625

Bai J, Ng S (2006) Confidence intervals for diffusion index forecasts and inference for factor-augmented regressions. Econometrica 74:1133–1150

Baillie RT (1996) Long memory processes and fractional integration in econometrics. J Econ 73:5–59

Bansal R, Kiku D, Yaron A (2016) Risks for the long run: estimation with time aggregation. J Monet Econ 82:52–69

Box GE, Jenkins GM, Reinsel GC, Ljung GM (2015) Time series analysis: forecasting and control. Wiley, London

Breitung J, Knüppel M (2018) How far can we forecast? Statistical tests of the predictive content. Deutsche Bundesbank Discussion Paper No. 07/2018

Carlstein E (1986) The use of subseries values for estimating the variance of a general statistic from a stationary sequence. Ann Stat 14:1171–1179

Chatfield C (1993) Calculating interval forecasts. J Bus Econ Stat 11(2):121–135

Cheng X, Liao Z, Schorfheide F (2016) Shrinkage estimation of high-dimensional factor models with structural instabilities. Rev Econ Stud 83(4):1511–1543

Christoffersen PF, Diebold FX (1998) Cointegration and long-horizon forecasting. J Bus Econ Stat 16:450–458

Clements MP, Taylor N (2003) Evaluating interval forecasts of high-frequency financial data. Appl Econ 18:445–456

Dehling H, Fried R, Shapirov O, Vogel D, Wornowizki M (2013) Estimation of the variance of partial sums of dependent processes. Stat Probab Lett 83(1):141–147

Diebold FX, Linder P (1996) Fractional integration and interval prediction. Econ Lett 50:305–313

Diebold FX, Rudebusch GD (1989) Long memory and persistence in aggregate output. J Monet Econ 24:189–209

Elliott G, Gargano A, Timmermann A (2013) Complete subset regressions. J Econ 177(2):357–373

Elliott G, Mller UK, Watson MW (2015) Nearly optimal tests when a nuisance parameter is present under the null hypothesis. Econometrica 83(2):771–811

Falk M (1984) Relative deficiency of kernel type estimators of quantiles. Ann Stat 12(1):261–268

Falk M (1985) Asymptotic normality of the kernel quantile estimator. Ann Stat 13(1):428–433

Ghalanos A (2017) rugarch: Univariate GARCH models. R package version 1.3-8

Gneiting T, Raftery AE (2007) Strictly proper scoring rules, prediction, and estimation. J Am Stat Assoc 102(477):359–378

Gonçalves S, de Jong R (2003) Consistency of the stationary bootstrap under weak moment conditions. Econ Lett 81(2):273–278

Han H, Linton O, Oka T, Whang Y-J (2016) The cross-quantilogram: measuring quantile dependence and testing directional predictability between time series. J Econ 193(1):251–270

Kim H, Swanson N (2014) Forecasting financial and macroeconomic variables using data reduction methods: New empirical evidence. J Econ 178:352–367

Kitsul Y, Wright J (2013) The economics of options-implied inflation probability density functions. J Financ Econ 110:696–711

Koenker R (2005) Quantile regression. Cambridge University Press, Cambridge

Künsch HR (1989) The jackknife and the bootstrap for general stationary observations. Ann Stat 17(3):1217–1241

Lahiri SN (2013) Resampling methods for dependent data. Springer, Berlin

Lütkepohl H (2006) Forecasting with varma models. In: Granger G, Granger WJ, Timmermann AG (eds) Handbook of economic forecasting, vol 1. Elsevier B.V., North Holland, pp 287–325

Marcellino M (1999) Some consequences of temporal aggregation in empirical analysis. J Bus Econ Stat 17(1):129–136

Müller U, Watson M (2016) Measuring uncertainty about long-run predictions. Rev Econ Stud 83(4):1711–1740

Pascual L, Romo J, Ruiz E (2004) Bootstrap predictive inference for arima processes. J Time Ser Anal 25(4):449–465

Pascual L, Romo J, Ruiz E (2006) Bootstrap prediction for returns and volatilities in garch models. Comput Stat Data Anal 50(9):2293–2312

Pastor L, Stambaugh RF (2012) Are stocks really less volatile in the long run. J Finance 67(2):431–478

Patton A, Politis DN, White H (2009) Correction to automatic block-length selection for the dependent bootstrap. Econ Rev 28(4):372–375

Politis DN, Romano JP (1994) The stationary bootstrap. J Am Stat Assoc 89:1303–1313

Politis DN, White H (2004) Automatic block-length selection for the dependent bootstrap. Econ Rev 23(1):53–70

Sheather SJ, Marron JS (1990) Kernel quantile estimators. J Am Stat Assoc 85:410–416

Silverman B (1986) Density estimation for statistics and data analysis. Chapman & Hall/CRC, London

Stock J, Watson M (2012) Generalised shrinkage methods for forecasting using many predictors. J Bus Econ Stat 30(4):482–493

Stock JH, Watson MW (2005) Understanding changes in international business cycle dynamics. J Eur Econ Assoc 3:968–1006

Sun S, Lahiri SN (2006) Bootstrapping the sample quantile of a weakly dependent sequence. Sankhyā Indian J Stat 68:130–166

White H (2000) A reality check for data snooping. Econometrica 68(5):1097–1126

Wu WB (2005) Nonlinear system theory: another look at dependence. Proc Natl Acad Sci USA 102(40):14150–14154 (electronic)

Wu WB, Woodroofe M (2004) Martingale approximations for sums of stationary processes. Ann Probab 32(2):1674–1690

Zhang T, Wu WB (2015) Time-varying nonlinear regression models: nonparametric estimation and model selection. Ann Stat 43(2):741–768

Zhou Z, Xu Z, Wu WB (2010) Long-term prediction intervals of time series. IEEE Trans Inform Theory 56(3):1436–1446